# Screening piezoelectricity in determination of flexoelectric coefficient at nanoscale

Yingzhuo Lun [a], Hao Zhou [b], Di Yao [a], Xueyun Wang [a] and Jiawang Hong [a,*]

[a] School of Aerospace Engineering, Beijing Institute of Technology, Beijing, P.R. China, 100081

[b] Beijing Key Laboratory of Intelligent Space Robotic Systems Technology and Applications, Beijing Institute of Spacecraft System Engineering, China Academy of Space Technology, Beijing, P.R. China,100094

[*]Corresponding author, E-mail: hongjw@bit.edu.cn; Tel.: 010-68915917

Piezoelectricity usually accompanies with flexoelectricity in polar materials which is the linear response of polarization to a strain gradient. Therefore, it is hard to eliminate piezoelectric effect in determination of pure flexoelectric response. In this work, we propose an analytical method to characterize the flexoelectric coefficient quantitatively at nanoscale in piezoelectric materials by screening piezoelectricity. Our results show that the flexoelectricity reduces the nanopillar stiffness while the piezoelectricity enhances it. With careful design of the shape of the nanopillars and measuring their stiffness difference, the flexoelectric coefficient can be obtained with the piezoelectric contribution eliminated completely. This approach avoids the measurement of electrical properties with dynamic load, which helps to reduce the challenge of flexoelectric measurement at nanoscale. Our work will be beneficial to quantitative characterization of flexoelectric properties and design of flexoelectric devices at nanoscale.

1. Introduction

Flexoelectricity is a novel electromechanical coupling effect which describes the linear coupling between the electric polarization and mechanical stress/strain gradient(Hong and Vanderbilt, 2011, 2013; Zubko et al., 2013; Wang et al., 2019). Different from the piezoelectricity existing only in the non-centrosymmetric materials, the flexoelectricity widely exists in all dielectric materials due to the strain gradient breaking the inversion symmetry of crystals. Therefore, the flexoelectricity provides potential opportunities for non-centrosymmetric materials in the applications of electromechanical systems(Chu et al., 2009; Jiang et al., 2013; Deng et al., 2014; Bhaskar et al., 2016; Zhang et al., 2017; Baroudi et al., 2018). However, the flexoelectric effect is relatively weak in bulk materials compared with the piezoelectric effect, it received little attention in past decades. Recently, the large flexoelectric coefficients were measured in the ferroelectric materials with high dielectric permittivity(Ma and Cross, 2005, 2003, 2002, 2001). In addition, the strain gradient can increase by 6-7 orders of magnitude compared with that in bulk materials if the size reduces to the nanoscale. Therefore, the flexoelectric effect becomes significantly enhanced and plays a critical role in various physical properties at nanoscale.

In order to investigate flexoelectric effect at nanoscale, it is necessary to obtain an accurate flexoelectric coefficient, which quantitatively describes the linear coupling between the polarization and strain/stress gradient. Until now, various direct and indirect experimental methods have been proposed to measure the flexoelectric coefficients of bulk materials, such as the cantilever bending method, three-point bending method(Zubko et al., 2007, 2008; Zhang et al., 2019), a four-point bending method(Ma and Cross, 2003), the pyramid compression method(Cross, 2006), as well as converse flexoelectric effect(Fu et al., 2006; Shu et al., 2014; Abdollahi et al., 2019) etc. Some novel techniques, such as split Hopkinson pressure bar method(Hu et al., 2018), shock wave method (Hu et al., 2017) and photorefractive method(Shandarov et al., 2012) were also proposed to measure bulk flexoelectric coefficients. At nanoscale, the flexoelectric coefficients can be obtained by measuring the atoms spacing and shift(Gao et al., 2018) or the mechanical behaviors(Zhou et al., 2016), or using nanoindentation techniques(Gharbi et al., 2011; Robinson et al., 2012; Hadjesfandiari, 2013).

Though there are many methods proposed to measure the flexoelectric coefficients, there still exist several issues that need to be solved. Firstly, the flexoelectric coefficient measured by above methods is effective coefficient which is a coupling of a multiple tensor components of flexoelectric coefficient. Secondly, one usually applies a dynamic mechanical load and measures the current or voltage generated by the flexoelectric effect. This measurement is challenge to perform at nanoscale. More importantly, the measured current or voltage contributes both from piezoelectric and flexoelectric effects, making it very challenge to exclude the piezoelectric contribution during the flexoelectric coefficient measurement. Though the measurement could be performed above Curie temperature to dismiss the bulk piezoelectric effect, the piezoelectricity in small polar regions(Narvaez and Catalan, 2014) and the surface piezoelectricity above Curie temperature(Dai et al., 2011; Narvaez et al., 2015; Zhang et al., 2018) may still contribute to the flexoelectric measurements. Therefore, it is necessary to develop a method which can eliminate the influence of piezoelectricity in flexoelectric coefficient measurement even below Curie temperature.

In this paper, we propose a method to exclude the piezoelectric contribution to the flexoelectric measurement at nanoscale. We develop a phenomenological model considered the piezoelectricity and flexoelectricity to investigate the nano-compression behavior of two different cross-section nanopillars. An analytical framework is established for eliminating the influence of piezoelectric effect in flexoelectric coefficient measurement at nanoscale. By rationally designing the dimensions of the nanopillars, a pure flexoelectric coefficient, excluding the contribution of piezoelectricity completely, can be obtained by measuring stiffness difference of two kinds of cross-section nanopillars. In addition, this method avoids measuring electric current with dynamic mechanical load, which reduces challenge of flexoelectric measurement at nanoscale. Our method provides a promising method to exclude piezoelectricity in the flexoelectric measurement at nanoscale, which could be beneficial to obtain accurate flexoelectric coefficient and design of flexoelectric nanodevices.

**2. Mechanical method for quantitatively characterizing the flexoelectric coupling coefficient**

In order to study the flexoelectric effect induced by stress gradient in dielectric material at nanoscale, a variable cross-section nanopillar is designed, as shown in Fig. 1 (a). The stress and its gradient along the height direction are induced under the axially compressive load, which excites the piezoelectric and flexoelectric effects.

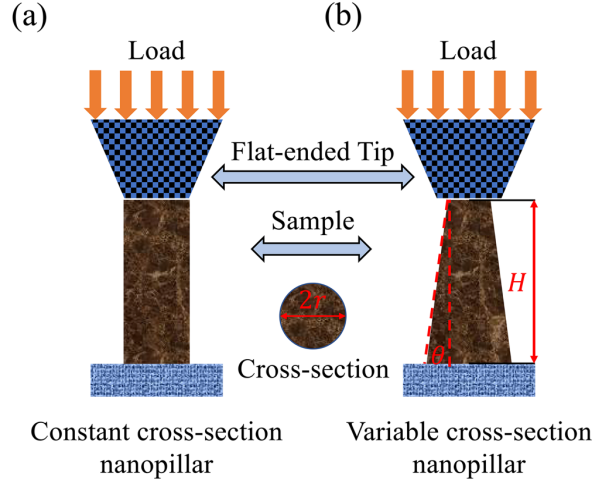

**Fig 1.** Schematic of nano-compression. (a) Variable cross-section nanopillar. (b) Constant cross-section nanopillar.

Since the nano-compression behavior of nanopillar can be approximately regarded as uniaxial compression along the height direction (i.e. $z$-axis), the theoretical model can be simplified to a one-dimensional model. Thus, the subscript will be omitted in the following derivation for the simplification. To describe the nano-compression behavior of the variable cross-section nanopillar, the thermodynamic potential density theory considering the piezoelectric and flexoelectric effect is employed. Taking the electric field $E$ and stress $\sigma$ as independent variables, the Gibbs energy $G$ can be expressed as Eq. (1) (Shen and Hu, 2010; Tichý, 2010)

$$G = \frac{1}{2}kE^2 + \frac{1}{2}s\sigma^2 + d\sigma E + \frac{1}{2}f\left[E\frac{d\sigma}{dz} - \sigma\frac{dE}{dz}\right] - DE - \sigma\varepsilon \tag{1}$$

where $E$ and $D$ are the electric field and the electric displacement, respectively; $\sigma$ and $\varepsilon$ are the stress tensor and the strain tensor; $k, s, d$ and $f$ are the second-rank dielectric permittivity tensor, the fourth-rank elastic compliance tensor, the third-rank piezoelectric tensor, and the fourth-rank flexoelectric coupling tensor, respectively. Particularly, the flexoelectric coupling tensor $f$ describes the coupling between the electric polarization and the stress gradient here. s

According to the Euler equation (i.e., $G_X - \frac{d}{dz}G_{X'} = 0$, where $X$ stands for $E$ or $\sigma$), the linear electromechanical constitutive equations for general piezoelectric materials can be obtained by minimizing total potential energy:

$$D(z) = kE(z) + d\sigma(z) + f\frac{d\sigma(z)}{dz} \qquad (2)$$

$$\varepsilon(z) = s\sigma(z) + dE(z) - f\frac{dE(z)}{dz} \qquad (3)$$

For the electric boundary condition, the electrical short circuit condition requires bottom electrodes and short circuit for the nanopillars, which is challenge for such pretreatments at nanoscale. Here, we choose the electrical open circuit condition where the nanopillars can be directly compressed without any electrode pretreatments. Then, the electric displacement satisfies the following Eq. (4), according to Gauss's law:

$$D_{z,z} = 0 \qquad (4)$$

Substituting Eq. (2) into Eq. (4), the electric field gradient can be obtained as:

$$\frac{dE(z)}{dz} = -\frac{d}{k}\frac{d\sigma(z)}{dz} - \frac{f}{k}\frac{d^2\sigma(z)}{dz^2} \qquad (5)$$

The electric field inside the nanopillars is contributed from the piezoelectric effect and flexoelectric effect:

$$E(z) = -\frac{d}{k}\sigma(z) - \frac{f}{k}\frac{d\sigma(z)}{dz} \qquad (6)$$

Substituting Eqs. (5) and (6) into Eq. (3), the strain along the height direction of nanopillar can be obtained as follows:

$$\varepsilon(z) = s\sigma(z) - \frac{d^2}{k}\sigma(z) + \frac{f^2}{k}\frac{d^2\sigma(z)}{dz^2} \tag{7}$$

It can be seen that the total mechanical strain consists of three parts: the first term of right side of Eq. (7) from the Hooke's law, the second term from the inverse piezoelectricity and the last term from the inverse flexoelectricity.

The stress along the height direction can be easily obtained according to the one-dimensional force balance equation:

$$F = \sigma(z) \cdot A(z) \tag{8}$$

where $F$ is the axially compressive load, $A(z)$ stands for the areas of cross-section at different height. Combining Eqs. (7) and (8), the total displacement $h$ along the height direction is obtained:

$$h = F\int_0^H \left[ (s - \frac{d^2}{k})A(z)^{-1} - \frac{\psi^2}{k}A(z)^{-2}A''(z) + 2\frac{f^2}{k}A(z)^{-3}[A'(z)]^2 \right]dz \tag{9}$$

where $H$ is the height of nanopillar, $A'$ and $A''$ are the first and second derivative of the areas of cross-section with respect to $z$.

For the variable cross-section nanopillar, the areas of cross-section $A_0(z)$ is related to the inclination angle $\theta$ and the radius $r_0$ of top surface:

$$A_0(z) = \pi r_0^2 (z \cdot \tan\theta / r_0 + 1)^2 \tag{10}$$

Substituting Eq. (10) into Eq. (9), the total displacement $h_0$ and compressive compliance $S_0$ of the variable cross-section nanopillar can be derived:

$$h_0 = F \cdot \left[ (s - \frac{d^2}{k})\frac{H}{\pi r_0^2}(\frac{\tan\theta}{r_0}H + 1)^{-1} + \frac{2f^2 \tan\theta}{k\pi r_0^3}[1 - (\frac{\tan\theta}{r_0}H + 1)^{-3}] \right] \tag{11}$$

$$S_0 = \frac{h_0}{F} = s\frac{H}{\pi r_0^2}(\frac{\tan\theta}{r_0}H + 1)^{-1} - \frac{d^2}{k}\frac{H}{\pi r_0^2}(\frac{\tan\theta}{r_0}H + 1)^{-1} + \frac{f^2}{k}\frac{2\tan\theta}{\pi r_0^3}[1 - (\frac{\tan\theta}{r_0}H + 1)^{-3}] \tag{12}$$

where the three terms of right side of Eq. (12) indicates the compressive compliance components related to the elastic property, the piezoelectric and flexoelectric effect, respectively.

Under the compressive stress field, the compressive compliance of the piezoelectric materials not only depends on its piezoelectricity but also flexoelectricity. Therefore, the flexoelectric coupling coefficient could be obtained from the mechanical property. Interestingly, the compressive compliances component has a quadratic relationship with the piezoelectric coefficient as well as the flexoelectric coupling coefficient. This means the influence of the piezo/flexoelectricity on the compressive compliance are independent of the sign (positive or negative) of its coefficients. Furthermore, it's worth noting that the sign of the second term and the third term is opposite, indicating that the piezoelectric and flexoelectric effects play competitive roles in the mechanical properties, i.e., piezoelectricity reduces the compressive compliance while flexoelectricity enhances it. This will be discussed in more details later.

Experimentally, the compressive compliance of nanopillar can be measured from the slope of displacement-load curve, which can be obtained from the nano-compression technique. Therefore, the flexoelectric coupling coefficient could be measured from this technique if the geometrical and physical properties of the nanopillar are obtained:

$$f = \pm \sqrt{\frac{S_0 - (s - \frac{d^2}{k})\frac{H}{\pi r_0^2}(\frac{\tan\theta}{r_0}H + 1)^{-1}}{\frac{2\tan\theta}{k\pi r_0^3}[1 - (\frac{\tan\theta}{r_0}H + 1)^{-3}]}} \tag{13}$$

This mechanical method not only avoids applying dynamical loading but also avoids measuring electric current measurement which requires the nanofabrication of electrodes and wire bonding at nanoscale. It only needs to perform nano-compression and measure the compressive compliance of nanopillars, which is much easier than the traditional flexoelectricity measurement with dynamical loading and electric current measurement. In order to obtain accurate flexoelectric coupling coefficient, it can be seen from Eq. (13) that the geometrical parameters and physical parameters need to be measured. The former could be precisely measured from Scanning Electronic Microscopy or Atomic Force Microscope, while the elastic compliance and piezoelectric coefficient (i.e. *s* and *d*) of nanoscale materials maybe quite different from their bulk properties(Liang et al., 2005; Zhu et al., 2012; Huan et al., 2014; Zheng et al., 2012; Banerjee et al., 2018). Therefore, it is of importance to dismiss the uncertainty of these physical parameters at nanoscale to obtain reliable flexoelectric coupling coefficient at nanoscale.

For the sake of avoiding measuring above physical parameters at nanoscale, as well as excluding the influence of piezoelectricity, we propose to introduce a constant cross-section nanopillar as a reference specimen, as shown in Fig. 1 (b), with the same height as the variable cross-section nanopillar. Then it is easy to get the area of the constant cross-section nanopillar:

$$A_1(z) = \pi r_1^2 \tag{14}$$

where $r_1$ is the radius of constant cross-section nanopillar. Substituting Eq. (14) into Eq. (9), the compressive compliance $S_1$ of constant cross-section nanopillar can be obtained as follows:

$$S_1 = s\frac{H}{\pi r_1^2} - \frac{d^2}{k}\frac{H}{\pi r_1^2} \tag{15}$$

where the first term is from the elastic property and the second term is from the piezoelectric effect. This is no flexoelectric contribution due to the homogenous stress distribution in the constant cross-section nanopillar.

Then, combining Eq. (12) and Eq. (15), the difference between the compressive compliances of two kinds of nanopillars can be written as follow:

$$S_0 - S_1 = (s - \frac{d^2}{k})\frac{H}{\pi}[\frac{1}{r_0^2}(\frac{\tan\theta}{r_0}H+1)^{-1} - \frac{1}{r_1^2}] + \frac{2\psi^2 \tan\theta}{k\pi r_0^3}[1-(\frac{\tan\theta}{r_0}H+1)^{-3}] \quad (16)$$

It can be seen that the first term of right side of Eq. (16) is from elastic and piezoelectric contribution, while the second term is purely from the flexoelectric contribution. If the radius of two kinds of nanopillars satisfy the following geometrical criterion:

$$r_1 = r_0\sqrt{\frac{\tan\theta}{r_0}H+1} \quad (17)$$

the Eq. (16) can be reduced to:

$$S_0 - S_1 = \frac{2f^2 \tan\theta}{k\pi r_0^3}[1-(\frac{\tan\theta}{r_0}H+1)^{-3}] \quad (18)$$

In this case, the difference of compressive compliances between variable cross-section and constant cross-section nanopillars only depends on the flexoelectric effect. Rewriting the Eq. (18), we can obtain the flexocoupling coefficient analytically as:

$$f = \pm\sqrt{\frac{(S_0-S_1)k\pi r_0^3}{2\tan\theta[1-(\frac{\tan\theta}{r_0}H+1)^{-3}]}} \quad (19)$$

It shows that the flexoelectric coupling coefficient can be determined analytically by the difference of the compressive compliances of variable and constant cross-section nanopillars. By employing an additional constant cross-section nanopillar, the piezoelectric contribution to the flexoelectric measurement will disappear completely if the constant cross-section nanopillar is selected carefully with its radius satisfying Eq. (17). What's more, the flexoelectric coupling coefficient can be obtained without measuring the elastic compliance and piezoelectric coefficients of materials at nanoscale. Both coefficients, which are different from their bulk properties and difficult to measure at nanoscale, can be eliminated completely after employing a constant cross-section specimen. Therefore, this analytical model can eliminate effectively the uncertainty of elasticity and piezoelectricity at nanoscale, and obtain a pure flexoelectric coupling coefficient excluding piezoelectric effect, which may make flexoelectric measurement more reliable at nanoscale.

## 3. Numerical results and discussions

In this section, we will perform numerical simulations based on the above model to investigate in details the competitive roles of piezoelectric and flexoelectric effects on the mechanical properties of nanopillars, as well as how the experimental errors affect the accuracy of flexoelectric coefficients. We choose typical ferroelectric BaTiO$_3$ single crystal nanopillars ($z$-axis along [001]) as an example, and the corresponding geometrical and physical parameters are listed in the Table I (Berlincourt and Jaffe, 1958; Hong et al., 2010; Narvaez et al., 2015). The $z$ axis points to the bottom surface with the origin located at the top surface (Fig. 2).

**Table I**. The geometrical and physical parameters of BaTiO$_3$ single crystal nanopillars at room temperature.

|       | $H$ (nm) | $H/r_0$ | $\theta$ (deg) | $s_{zz}$ ($10^{-12}$ m$^2$N$^{-1}$) | $k_{zz}$ ($\varepsilon_0$) | $d_{zz}$ (pC/N) | $\mu_{zz}$ (μC/m) |
|-------|----------|---------|----------------|-------------------------------------|---------------------------|-----------------|-------------------|
| value | 300      | 5       | 10             | 15.7                                | 168                       | 85.6            | 1.0               |

To show the nano-compressive state of a variable cross-section nanopillar with piezoelectricity and flexoelectricity, the distribution of compressive strain along the height direction of the nanopillars with different height are displayed in Fig 2. It can be seen that the type II curves with piezoelectric effect are below the type I curves with sole elasticity, while the type III curves with flexoelectric effect are beyond type I curves. This suggests the piezoelectricity and flexoelectricity play competitive roles on the compressive compliance and the former reduces the compliance while the latter enhances it. If both piezoelectric and flexoelectric effects are considered, the strain distribution curve (type IV) can be below (Fig. 2 (a)), beyond (Fig. 2 (c)) or crossing with (Fig. 2 (b)) the curve I, depending on the height of the nanopillars. In nanopillar with 500nm height, the strain distribution curve IV is below type I curve, suggesting the piezoelectricity plays a more significant role than that of flexoelectricity (Fig. 2 (a)), When the height decreases to 100nm, the strain distribution curve IV is far beyond curve I and very close to curve III with flexoelectric effect (Fig. 2 (c)). This indicates the flexoelectricity plays a dominant role in this smaller nanopillar with higher strain gradient, in which the piezoelectric effect is negligible. With proper dimension size of nanopillar, the effect of piezoelectricity and flexoelectricity could be canceled out and the strain distribution is close to pure elastic curve I, as shown in Fig. 2 (b).

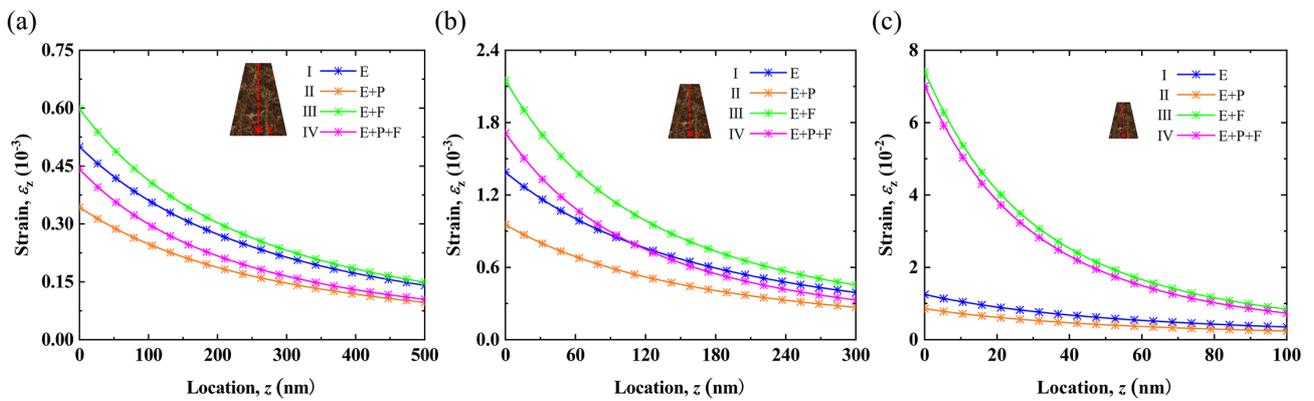

**Fig 2.** The distribution of compressive strain along the height direction of the variable cross-section BaTiO$_3$ single crystal nanopillar subjected to mechanical uniaxial compressive load (1 μN). The height of the nanopillar is (a) 500 nm, (b) 300 nm, (c) 100 nm, respectively. Four different types of compressive strain distributions shown in plots are obtained with the considerations of (I) pure elasticity (E); (II) elasticity (E) and piezoelectricity (P); (III) elasticity (E) and flexoelectricity (F); and (IV) elasticity (E), piezoelectricity (P) and flexoelectricity (F).

To further understand the influence of piezoelectric and flexoelectric effect on the mechanical properties, the displacement-load curves of the variable cross-section nanopillars are plotted in Fig. 3 (a). Compared to the compressive compliance (slope of the displacement-load curves) contributed from elasticity alone, the compressive compliance decreases by 34 % with the piezoelectric effect considered, while it increases by 33 % if the flexoelectric effect is taken into account. This shows that the flexoelectric effect gives rise to stiffness softening while the piezoelectric effect causes stiffness hardening, in line with previous report about piezoelectric toughening(Tichý, 2010). When both piezoelectric and flexoelectric effect are considered, their effects nearly cancel out and the compressive compliance increases slightly by 1.7 %. In addition, as the height decreases, the stiffness softening becomes more significant and the piezoelectric effect is negligible because the flexoelectric effect plays dominating role in small nanopillars which has higher strain gradient, as shown in Fig. 3 (b). There exists a critical height (309 nm for the nanopillar considered here) below which the flexoelectricity plays a dominating role to soften stiffness. Beyond this critical height, the piezoelectricity will have more significant role to make nanopillar hardening.

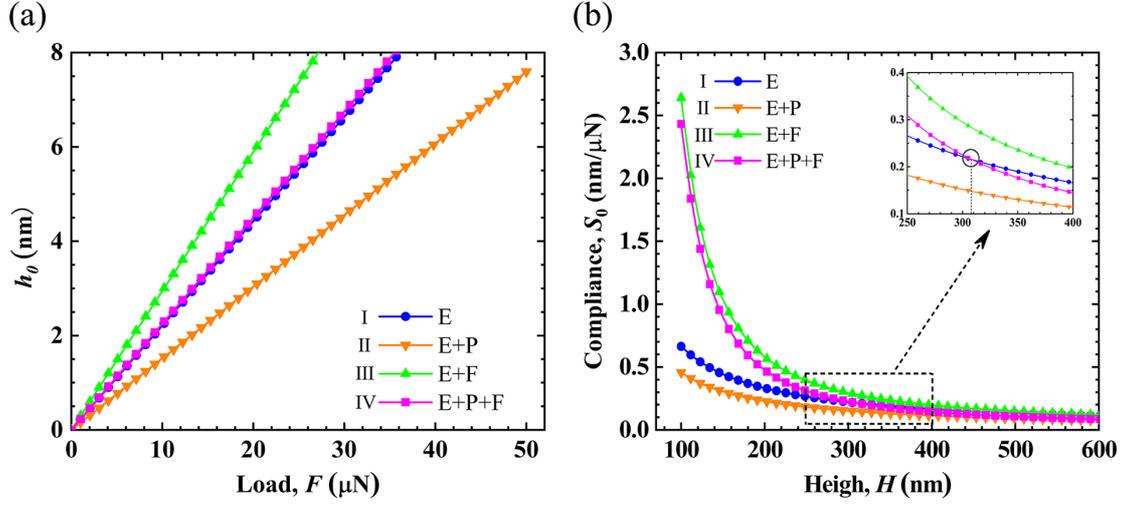

**Fig 3.** (a) Displacement-load curves of the variable cross-section nanopillars. $H = 300$ nm, $H/r_0 = 5$, $\theta = 10°$. (b) The compressive compliances *vs.* height of nanopillars with piezoelectric and flexoelectric effects. $H/r_0 = 5$, $\theta = 10°$.

For the purpose of quantifying the flexoelectric stiffness softening effect, we define a stiffness softening factor $\delta = (S_0-S_1)/S_0$, where $S_0$ and $S_1$ are the compressive compliance of the variable and constant cross section nanopillar above-mentioned, respectively. The two nanopillars are of equal height and the geometric dimensions satisfy the geometrical criteria of Eq. (17). It is expected that the stronger flexoelectric effect, i.e., the higher flexoelectric coefficient or larger strain gradient, the more significant stiffness softening. Indeed, as can be seen from Fig. 4 (a), a shorter nanopillar with a higher flexoelectric coefficient possesses more remarkable stiffness softening. For example, nanopillars with 100 nm height and flexoelectric coefficient of 0.5 μC/m induces more than 50% stiffness drop, while the stiffness softening is negligible for the nanopillars with flexoelectric coefficient less than 0.05 μC/m. The geometry of samples also has important effect on the stiffness softening, as shown in Fig.

4 (b). With the inclination angle increases, the cross-section area changes more rapidly in the height direction of nanopillar and induces larger strain gradient. Higher aspect ratio nanopillars also possesses larger strain gradient. As a result, a larger flexoelectric effect is produced and the stiffness softening becomes more significant.

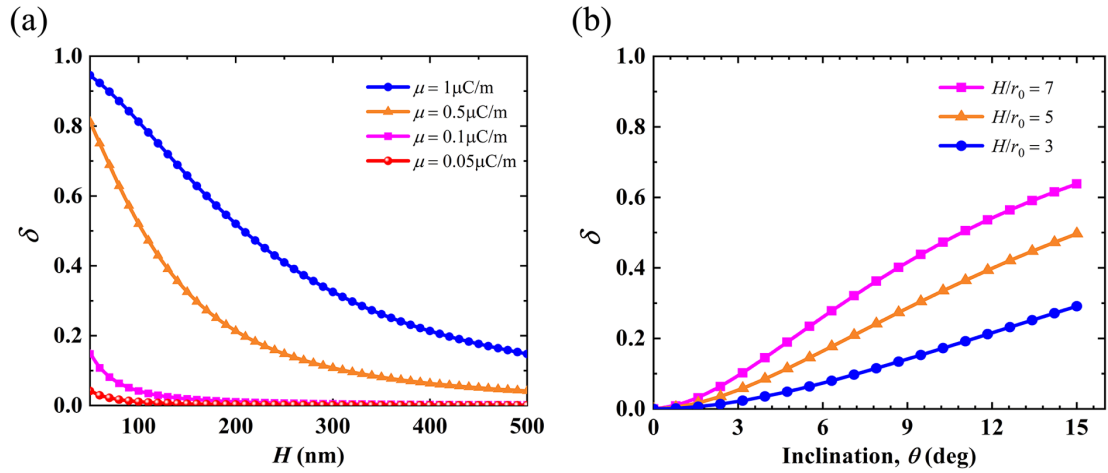

**Fig 4.** Stiffness softening of nanopillars. (a) Influence of height and flexoelectric coefficient. $H/r_0 =$ 5, $\theta = 10°$. (b) Influence of inclination angle and aspect ratio. $\mu = 1.0$ μC/m, $H = 300$ nm.

According to the theoretical model developed in pervious section, the influence of the piezoelectricity can be eliminated completely in principle and the pure flexoelectric coupling coefficient can be obtained if the geometrical criterion Eq. (17) is satisfied. We propose to fabricate the variable cross section nanopillar first and measure its geometrical dimensions precisely. And then, the constant cross-section nanopillar is manufactured with the radius determined by Eq. (17). However, in practice, after fabrication, the radius of constant cross-section nanopillar may not exactly the same as the ideal radius from Eq. (17). This may bring some uncertainty to measure the flexoelectric coupling coefficient by using this method. We next analyze how this fabrication error affects the measurement and how to minimize this effect.

Fig. 5 shows that a larger (smaller) flexoelectric coupling coefficient than the true value will be obtained if the fabricated radius is larger (smaller) than the ideal radius. But this measurement error can be reduced effectively if introducing larger strain gradient by choosing shorter specimen or larger inclination angles. For example, for the specimen with height 300 nm, the practical radius deviating from the ideal size by 6 % will introduce 10.8 % measurement error of flexoelectric coefficient, but it reduces to only 1.3 % for the 100nm height specimen. Similarly, using specimen with large inclination angles will also help to reduce the measurement error of flexoelectric coefficient if the fabricated radius deviates from the ideal radius. In fact, this measurement error origins from the piezoelectric contribution if the fabricated radius is not equal to the ideal radius determined by Eq. (17). Therefore, in order to reduce this measurement error, one should choose the specimen with short height and/or large inclination angles to enhance the flexoelectric effect and weaken the piezoelectric effect in the measurement.

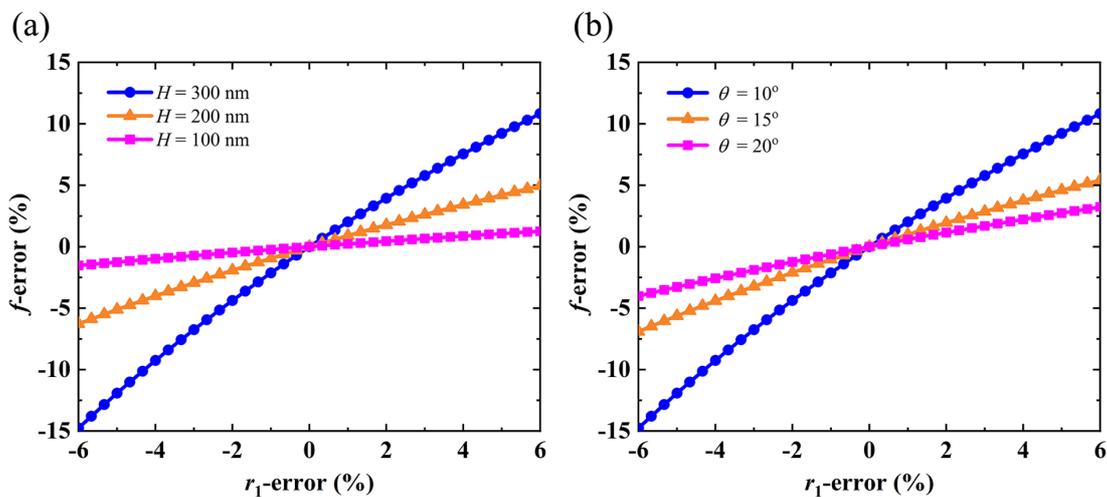

**Fig 5.** The measurement error of flexoelectric coupling coefficient introduced by the fabricated error of radius, (a) the influence of the height, $\theta = 10°$, (b) the influence of the inclination angles, $H = 300$ nm.

## 4. Conclusions

In this paper, we develop a phenomenological model considered the flexoelectricity and piezoelectricity to investigate the nano-compression behaviors of two different kinds of nanopillars, under the electrical open circuit condition. The results indicate that the piezoelectricity and flexoelectricity play a competitive role on the mechanical properties of nanopillars, in which the former causes stiffness hardening while the latter enhances give rise to stiffness softening. Based on this, we propose to measure pure flexoelectric coupling coefficient eliminating the piezoelectric contribution completely by measuring the compressive compliances of two different kinds of nanopillars with carefully designed shapes. Our results show that choosing a short specimen with a large inclination angle can reduce effectively the measurement error from the fabricated radius deviating the ideal radius. Our work provides a promising approach to measure flexoelectric coupling coefficient without piezoelectric contribution, and avoids measuring the electric current with dynamic load at nanoscale. This will help us to have better understanding of novel phenomena induced by ferroelectric at nanoscale and design flexoelectric nanodevices.

**Acknowledgments**

This work was supported by the National Natural Science Foundation of China (Grant No. 11572040, 11702019 and 11604011) and Beijing Natural Science Foundation (Grant No. Z190011). Y.Z.L. is supported by Graduate Technological Innovation Project of Beijing Institute of Technology (grant 2019CX20002). H.Z is supported by Young Elite Scientists Sponsorship Program by China Association for Science and Technology (Grant No. 2017QNRC001).